\documentclass[preprint,showpacs,preprintnumbers,amsmath,amssymb]{revtex4}

\usepackage{graphicx}
\usepackage{dcolumn}
\usepackage{bm}
\usepackage{afterpage}

\usepackage{color}

\begin{document}


\title{Critical Networks Exhibit Maximal Information Diversity in
Structure-Dynamics Relationships}

\author{Matti Nykter$^{1,2}$, Nathan D. Price$^{2,4}$, Antti Larjo$^1$, Tommi
Aho$^1$, Stuart A. Kauffman$^3$, Olli Yli-Harja$^1$,
Ilya Shmulevich$^{2,*}$}
\affiliation{%
$^1$Institute of Signal Processing, Tampere University of Technology, Tampere,
Finland \\
$^2$Institute for Systems Biology, Seattle, WA, USA \\
$^3$Institute for Biocomplexity and Informatics, University of Calgary,
Calgary, AB, Canada \\
$^4$Department of Chemical and Biomolecular Engineering, University of
Illinois, Urbana, IL, USA\\
$^*$ To whom correspondence should be addressed (ishmulevich@systemsbiology.org)
}%

\date{\today}

\pacs{Valid PACS appear here}

\begin{abstract}

Network structure strongly constrains the range of dynamic behaviors
available to a complex system.  These system dynamics can be classified
based on their response to perturbations over time into two distinct
regimes, ordered or chaotic, separated by a critical phase transition.
Numerous studies have shown that the most complex dynamics arise near the
critical regime.  Here we use an information theoretic approach to study
structure-dynamics relationships within a unified framework and show that
these relationships are most diverse in the critical regime.


\end{abstract}

\maketitle

The structural organization, or topology, of a complex system 
strongly constrains the range of dynamical behaviors available to
the system.
Understanding structure-dynamics relationships in networks is a major goal
of complex systems research \cite{nochomovitz06, variano04}. However,
general principles behind such 
relationships are still lacking, in part due to the lack of sufficiently
general formalisms for studying structure and dynamics within a common
framework. Numerous relationships between specific structural and dynamical
features of networks have been investigated \cite{aldana03b, albert00,
kauffman03, shmulevich03, kauffman04}.  For example, structure can be
studied by means of various graph-theoretic features of network topologies
such as degree distributions \cite{barabasi99} or modularity
\cite{watts98}, or in 
terms of classes of updating rules that generate the dynamics
\cite{shmulevich03, kauffman04}.
Aspects of 
dynamical behavior include transient and steady-state behavior and the
response of the system to perturbations 
\cite{aldana03b, kauffman93}. 

Relating structure to dynamics is important for understanding emergent
behaviors because the structure in a complex system directly affects
emergent properties such as robustness, adaptability, decision-making and
information processing \cite{klemm05, aldana03b}. 
An important aspect of many complex
dynamical systems is the existence of two dynamical regimes, ordered and
chaotic, with a critical phase transition boundary between the two
\cite{aldana03, derrida86, mestl97}.  These regimes
profoundly influence emergent dynamical behaviors, and can be observed in
different ensembles of network structures.  Networks operating in the
ordered regime are intrinsically robust, but exhibit simple dynamics.  This
robustness is reflected in the dynamical stability of the
network both under structural perturbations and transient
perturbations. 
Contrary to this, networks in the chaotic regime are
extremely sensitive to small perturbations, which rapidly propagate
throughout the entire system and hence fail to exhibit a natural basis for
robustness and homeostasis. The phase transition between the ordered and
chaotic regimes represents a tradeoff between the need for stability and the
need to have a wide range of dynamic behavior to respond to a variable
environment. It has long been hypothesized that biological networks operate
near this phase transition \cite{kauffman93}.  Recent evidence suggests that
biological networks are not chaotic \cite{shmulevich05, ramo06}.

A theme that has emerged in many contexts in systems theory is that complex
systems operating near the phase transition exhibit maximally ``interesting''
dynamics \cite{krawitz07, ramo07, luque00b, sole96, langton90, packard88}.
Complex coordination of information processing seems to be 
maximized near the phase transition.  Information theory provides a common
lens through which we can study both structure and dynamics of complex
systems within a unified framework. Indeed, since network structures as well as
their dynamic state trajectories are objects that can be represented on a
computer, the information encoded in both can be compared and related.
Unlike Shannon's information, which is defined in terms of distributions,
Kolmogorov complexity is a suitable framework 
for capturing the information embedded in individual objects of finite
length \cite{li97}.

Recent developments in information theory have demonstrated that Kolmogorov
complexity can be used to define an absolute information distance between
two objects \cite{bennett98, li04}, called herein universal information
distance.  This distance 
metric is universal in that it can be applied to any objects that can be
stored on a computer (e.g. networks or genome sequences), and uniquely
specifies an information distance without parameters of any kind.  Thus, it
is suitable for comparing the information content of two objects.  Although
this distance is, like the Kolmogorov complexity itself, uncomputable, it
can be approximated by real-world data compressors (herein, \texttt{gzip}
and \texttt{bzip2})
to yield the normalized compression distance (NCD) \cite{li04}, defined as
\begin{equation}
NCD(x,y)=\frac{C(xy)-\min\{C(x),C(y)\}}{\max\{C(x),C(y)\}},
\end{equation}
where $C(x)$ is the compressed size of $x$ and $xy$ is the
concatenation of the strings $x$ and $y$.  Cilibrasi and Vit{\'a}nyi
\cite{cilibrasi05}  have
demonstrated that the NCD can be used for clustering using a real-world
compressor with remarkable success, approximating the provable optimality of
the (theoretical) universal information distance.

Herein we apply the NCD to various classes of networks to study their
structure-dynamics relationships.  The proposed methods can be applied in
principle to any model class that can be represented on a computer.  We can
represent the state of a network by a set of $N$ discrete-valued variables, 
$\sigma_1,\sigma_2,{\ldots},\sigma_N$.
Examples include Boolean networks ($\sigma_n\in\{0,1\}$) \cite{aldana03,
kauffman93, kauffman69b}, ternary networks
($\sigma_n\in\{0,1,2\}$), and so forth,
depending on the level of detail desired.

To each node $\sigma_n$ we assign a set of $k_n$ nodes,
$\sigma_{n_1},\sigma_{n_2},{\ldots},\sigma_{n_{k_n}}$, which control the
value of $\sigma_{n}$
through the equation
$\sigma_n(t+1)=f_n(\sigma_{n_1}(t),\sigma_{n_2}(t),{\ldots},\sigma_{n_{k_n}}(t))$.
In the case of Boolean networks, we choose the 
functions $f_n$ randomly from the set of all possible Boolean functions such
that 
for each configuration of its $k_n$ arguments, $f_n=1$ with probability $p$,
known as the 
bias. The average of $k_n$, denoted by $K$, is called the average network
connectivity. Here, we consider two types of wiring of nodes: 1) random,
where each node has exactly $K$ inputs chosen randomly; and 2) regular, where
nodes are arranged on a regular grid such that each node takes inputs from
its $K$ neighbors. For random wiring the critical phase transition curve is
defined by $2Kp(1-p)=1$ \cite{derrida86}. Thus, for $p=0.5$ (unbiased) 
$K=1, 2,$ and $3$ correspond to ordered, critical, and chaotic regimes,
respectively.

Making use of the universal information distance as approximated by the
NCD, we are able to compare networks without reducing them to arbitrary sets
of features (e.g., graph properties). 
Indeed, the NCD uses the information and regularities embedded in the
network structure to pick up the relative differences in the structural
complexities of the networks. In order for NCD to most effectively capture
structural differences, a network structure representation that is most
amenable to compression (i.e. approximating the Kolmogorov complexity as
well as possible) should be used. For example, for Boolean networks, we
represent the connections by distances along an arbitrary linear arrangement
of the nodes, making the regularities in the network structure more easily
observable.
See supplementary material for details on the encoding of network structures.

\begin{figure}
\centering
\includegraphics[width=\textwidth]{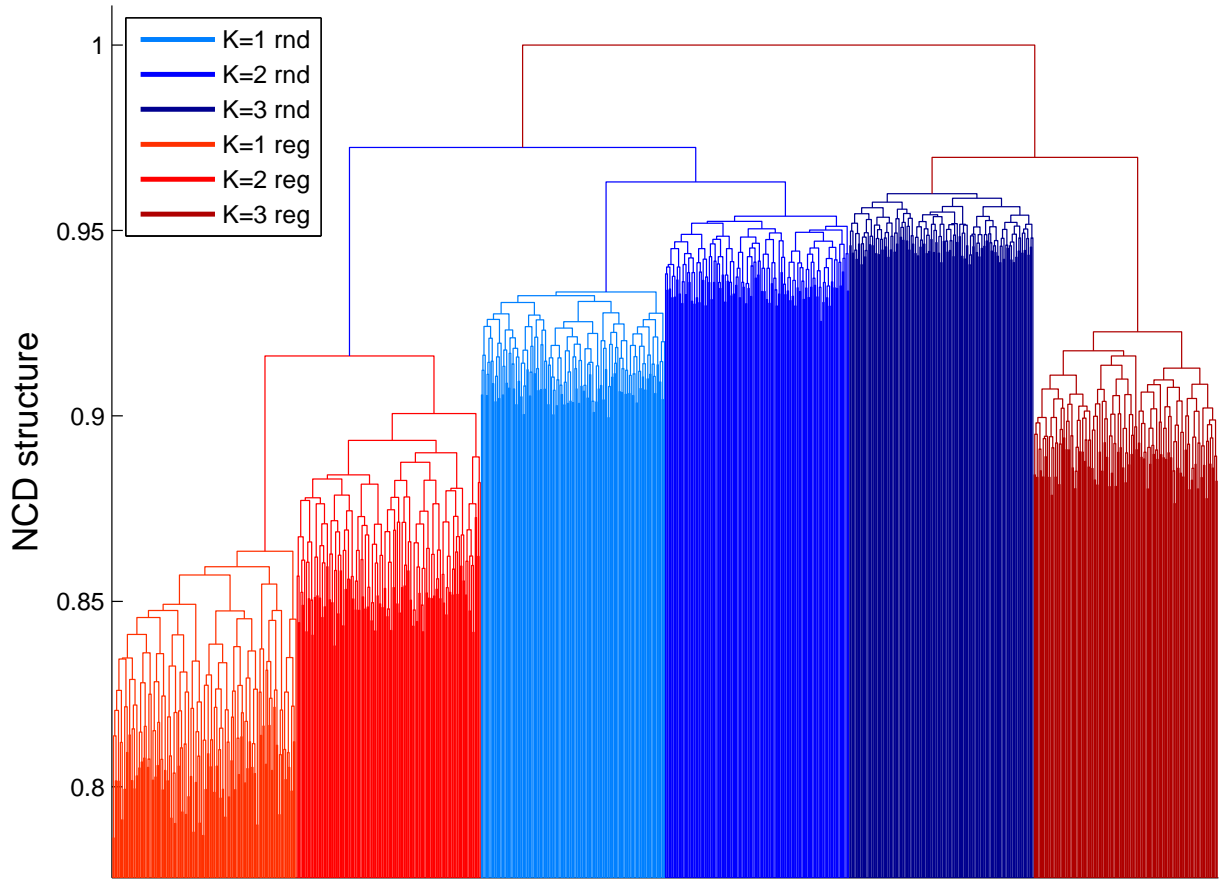}
\caption{Six ensembles of
random Boolean networks ($K = 1, 2, 3$ each with random or regular topology;
$N = 1000$) were used to generate 30 networks from each ensemble.
The normalized compression distance (NCD) was applied to all pairs of
networks. Hierarchical clustering with the complete linkage
method was used to build the dendrogram from the NCD distance matrix (see supplementary
material for details on clustering). 
Networks from different ensembles are clustered
together, indicating that
intra-ensemble distances are smaller than inter-ensemble distances.
}
\label{fig:structure}
\end{figure}

To illustrate this, we
generated six Boolean network ensembles ($N = 1000$) with two different wiring
topologies: random and regular, each with $K = 1, 2,$ or $3$. As Figure
\ref{fig:structure}
illustrates, all of the different ensembles considered are clearly
distinguishable. 

\begin{figure}
\centering
\includegraphics[height=0.8\textheight]{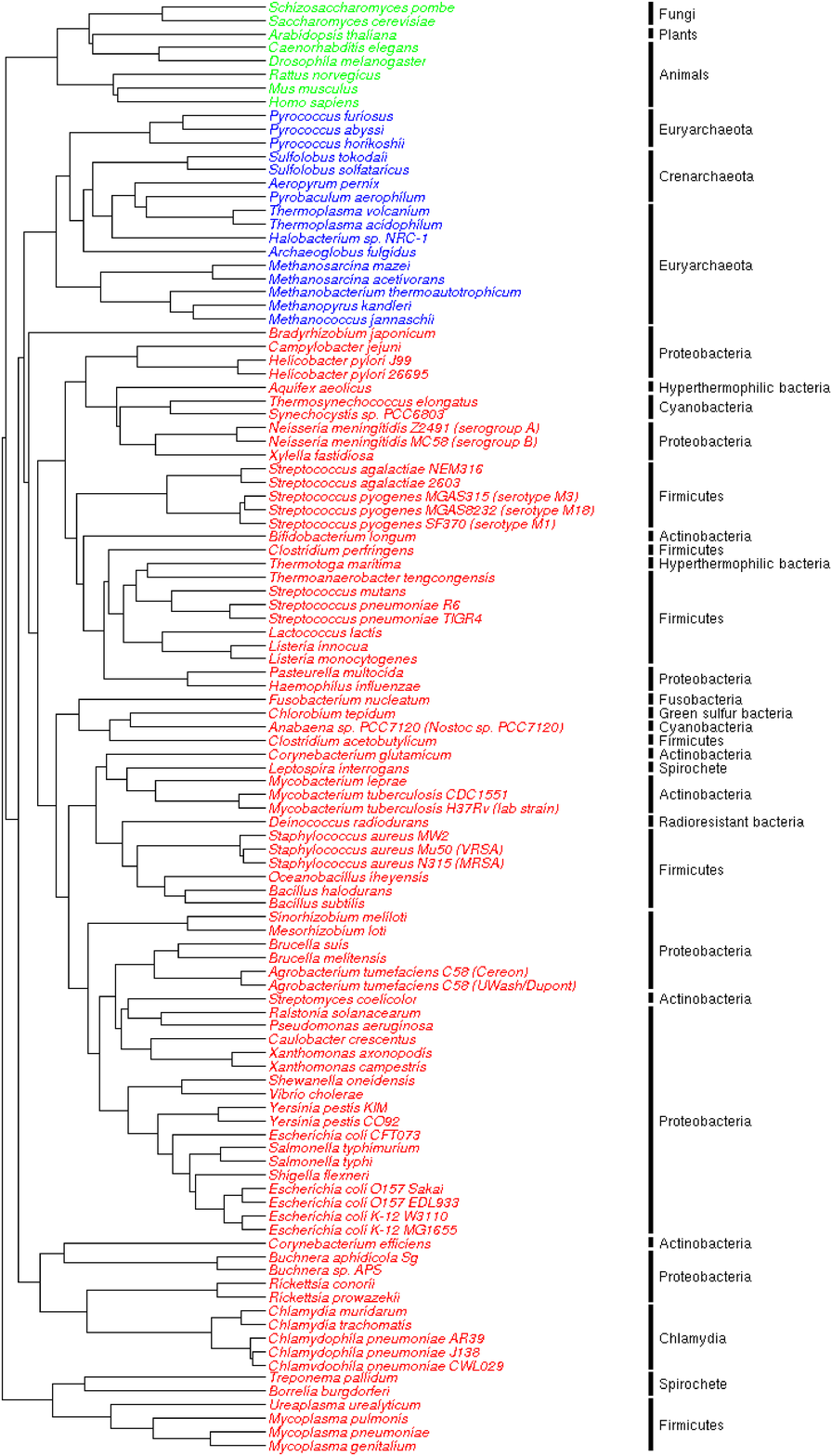}
\caption{A phylogenetic tree generated using NCD applied to all pairs of
metabolic network structures from 107 organisms in KEGG. 
Different domains of life appear in distinct branches. Bacteria are shown 
in red, archaea in blue, and eukaryotes in green. Subclasses of species within
each domain are labeled on
the right. Parasitic bacteria (bottom branch) are separated from the rest as
observed earlier. 
This separation of the domains of life indicates that the method is able to
discover the fundamental structural differences in the metabolic networks. 
}
\label{fig:phylogen}
\end{figure}

To demonstrate that the NCD is able to capture meaningful structural
relationships between networks, we applied it to compute the pairwise
distances between the metabolic networks of 107 organisms from the KEGG
database \cite{kanehisa00} (see supplementary material for details). The
resulting 
phylogenetic tree, generated using the complete linkage method, is shown in
Figure \ref{fig:phylogen}.  The organisms are clearly grouped into the three
domains of life. 
The archaea and eukaryotes both separate into distinct branches of
phylogenetic tree based on the
information content of their metabolic networks.  The bacteria form three
distinct branches, with parasitic bacteria encoding more limited metabolic
networks separating from the rest, as has been observed previously
\cite{podani01}. The
fact that the phylogenetic tree reproduces the known evolutionary
relationships suggests that the NCD successfully extracts structural
information embedded in networks.

\begin{figure}
\centering
\includegraphics[width=\textwidth]{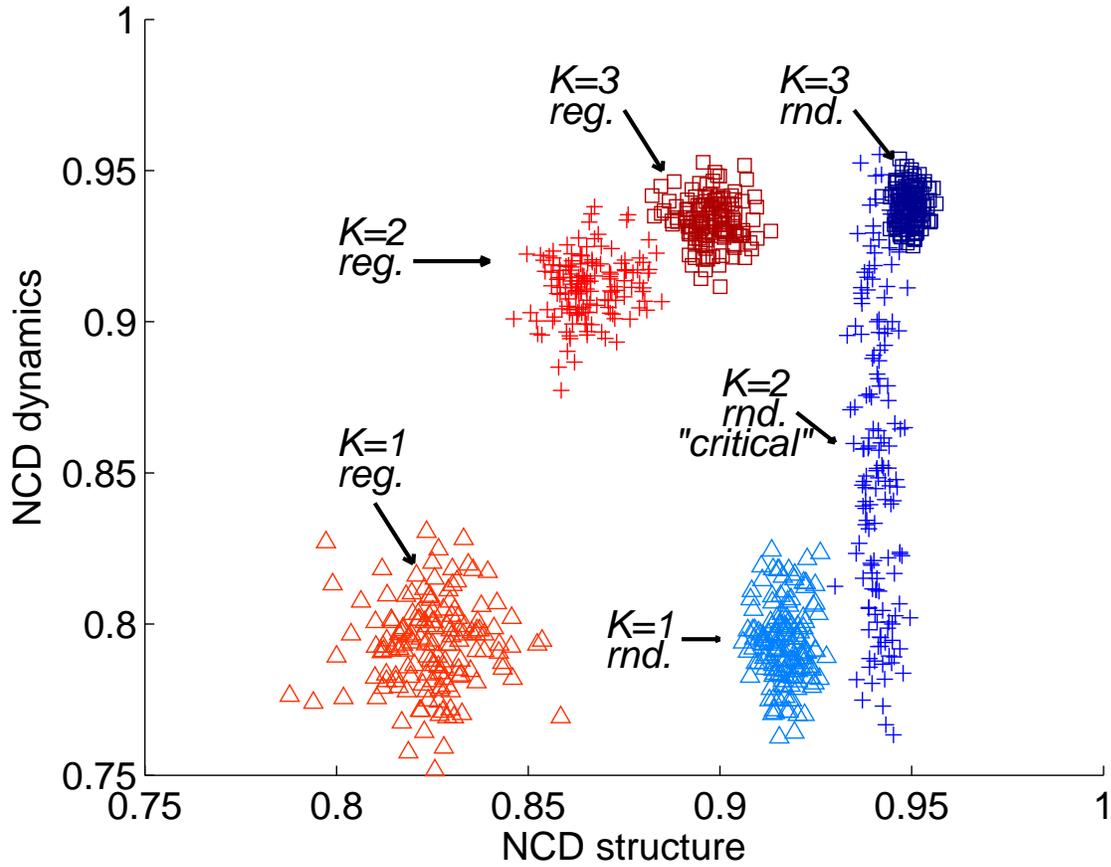}
\caption{The normalized compression distance (NCD) applied to network
structure and
dynamics. Six ensembles of random Boolean networks ($K = 1, 2, 3$ each with
random or regular topology; $N = 1000$) were used to generate 150 networks
from each ensemble. NCDs were computed between pairs of networks (both
chosen from the same ensemble) based on their structure (x-axis) and their
dynamic state trajectories (y-axis). Different ensembles are clearly 
distinguishable. The critical ensemble is more elongated, implying
diverse dynamical behavior. 
}
\label{fig:relation}
\end{figure}

We used NCD to study the relationship between structural information and
dynamical behavior within a common framework.  Within each of the above 6
ensembles, we generated $150$ networks and calculated the NCD
between all pairs of network structures and 
between their associated dynamic state trajectories.  After running the
network dynamics 100 steps from a random initial state to ensure that the
network is not in a transient state,
state trajectories were collected for 10 consecutive
time steps and the states were concatenated into one vector (see supplementary
material for details on the encoding of dynamics). 
As is the case with the structure of
the networks, NCD is expected to detect the regularities in the state
trajectories and thus uncover their intrinsic similarities.
The relationship between structure and dynamics was
visualized by plotting the structure-based NCD 
versus the dynamics-based NCD for pairs of networks within each ensemble
(Figure \ref{fig:relation}).  
All network ensembles were clearly distinguishable based on their
structural and dynamical information.  
It is also interesting to note that within-ensemble distances increase
as dynamical complexity increases. Similarly, as the structure gets more
complex, from regular to random wiring, or with increasing in-degree, the
within-ensemble distances increase.

The critical ensemble
($K = 2$, random wiring) exhibited a distribution that is markedly more
elongated along the dynamics axis as compared to the chaotic and ordered
ensembles, supporting the view that
critical systems exhibit maximimal
diversity.  The wide spread of points for the
critical network ensemble in Figure \ref{fig:relation} shows that their dynamics range
between those of ordered and chaotic ensembles.  Indeed, very different
network structures can yield both relatively similar and dissimilar
dynamics, thereby demonstrating the dynamic diversity exhibited in the
critical regime.  Thus, the universal information distance provides clear
evidence that the most complex relationships between structure and dynamics
occur in the critical regime.

In summary, we have demonstrated that NCD is a powerful tool for extracting
structure-dynamics relationships. It is fascinating that the analyses
presented herein are possible using only the compressibility of a file
encoding the network or its dynamics without needing to select any
particular network parameters or features. This approach allows us to
study, under a unified information theoretic framework, how a change in
structural complexity affects the dynamical behavior, 
or vice versa.

\section*{Acknowledgments}

This work was supported by NIH-GM070600 (IS, SAK), NIH-GM072855 (IS),
NIH-P50 GM076547 (IS), Academy of Finland (project No. 213462 OY-H, No.
120411 and No. 122973 MN), 
American Cancer Society 
PF-06-062-01-MGO (NDP). The authors are grateful to Leroy Hood, David Galas,
Jared Roach, and Alok Srivastava for useful discussions. 

%
%
%
%
%
%
\afterpage{\clearpage}

\afterpage{\clearpage}

\newpage

\bibliography{abrvs,paper}

\end{document}